\documentclass[amsmath, amssymb, aps, preprint, bibliography]{revtex4-1}
\usepackage{amssymb, amsmath, bm, graphicx,color}
\usepackage{epstopdf}
\usepackage{array}

\begin{document}


\title{Thermal Conductivity Enhancement by Surface Electromagnetic Waves Propagating along Multilayered Structures with Asymmetric Surrounding Media}


\author{Mikyung Lim$^1$, Jose Ordonez-Miranda$^2$, Seung S. Lee$^1$, Bong Jae Lee$^{1*}$, Sebastian Volz$^{3,4}$}
\email{bongjae.lee@kaist.ac.kr; volz@iis.u-tokyo.ac.jp}
\affiliation{$^1$Department of Mechanical Engineering, Korea Advanced Institute of Science and Technology, Daejeon 34141, South Korea\\
$^2$Institut Pprime, CNRS, Université de Poitiers, ISAE-ENSMA, F-86962 Futuroscope Chasseneuil, France\\
$^3$LIMMS/CNRS-IIS(UMI2820), Institute of Industrial Science, University of Tokyo, Tokyo 153-8505, Japan \\
$^4$Laboratoire d'Energétique Moléculaire et Macroscopique, Combustion, UPR CNRS 288 CentraleSupélec, Université Paris-Saclay, France}


\date{\today}

\begin{abstract}
Enhancement of thermal conductivity via surface electromagnetic waves (SEWs) supported in nanostructures has recently drawn attention as a remedy for issues raised due to the reduction of thermal conductivity in nanoscale confinement. Among them, multilayered structures on a substrate are prevalent in nano-sized systems, such as electronic nanodevices, meaning that analysis on those structures is indispensable. In this work, three basic multilayered structures are selected and the analytical expressions for SEWs supported in each structure are derived. This analytical approach enables us to figure out which factors are crucial for enhancing SEW thermal conductivity using multilayers. It is also found that the solution can be extended to various materials and provide the guidelines on which configurations are desirable for increasing the thermal conductivity. Furthermore, the analytical solutions reduce the calculation time significantly such that the optimal configuration, which can additionally yield SEW thermal conductivity of 1.27 W/m$\cdot$K corresponding to  90\% of the thermal conductivity of bulk glass, is found with the genetic algorithm. This study thus provides a new method for efficiently managing thermal issues in nano-sized devices.
\end{abstract}
\maketitle

\section{Introduction}
Thermal energy transport via surface electromagnetic waves (SEWs) propagating along thin films has been intensively investigated over the last decade \cite{chen2005surface,chen2010heat,ordonez2013anomalous,ordonez2014thermal,ordonez2014quantized,tranchant2015measurement} because of its potential application to compensate the reduction of the thermal performance of systems when their sizes are scaled down to nanoscales \cite{chen2010heat}. In particular, the booming miniaturization of electronic devices has been leading to a decrease of the size of their components to a few nanometers, which is smaller than the mean free path of the electrons or phonons of the constituting materials and reduces their effective thermal conductivities \cite{zhang2007nano}. Given that the problem of heat spreading can generate hot spots, which overheat the electronic nanodevices and deteriorate their performance \cite{semenov2006impact}, the additional energy transport mechanism driven by SEWs supported in nano-sized systems has been regarded as a promising remedy to tackle this issue of overheating \cite{chen2005surface,chen2010heat,ordonez2013anomalous,ordonez2014thermal,ordonez2014quantized,tranchant2015measurement}.

SEWs are bounded at the interface along which they propagate carrying energy \cite{raether1988surface,burke1986surface, yang1991long, kume1998long}. The propagation length of these SEWs can be drastically increased as the thickness of the suspended glass layer decreases \cite{chen2005surface,ordonez2013anomalous}; however, when this single glass layer is placed on a substrate (i.e., asymmetric surrounding media), the frequency interval where SEWs can be supported becomes severely restricted \cite{ordonez2013anomalous}, and thus, an additional energy transport via SEWs is rather limited. Furthermore, besides the single glass layer, a thermal conductivity enhancement by SEW within mutilayered structures should also be considered for wide-range of applications.

Although SEWs supported in periodic multilayered structures has been widely investigated \cite{dereux1988polaritons,mendialdua1994bulk,lim2018optimization, ben2010surface, biehs2013super,guo2012broadband, wang2018two}, their contribution on thermal conductivity has been rarely reported. A previous work \cite{ordonez2014thermal} explored the thermal conductivity enhancement of a suspended nano-layered system due to the propagation of SEWs by regarding nano-layers as a single layer with the effective permittivity. Because this effective permittivity only depends on a relative thickness of the constituting layers, it cannot fully take account of the configuration of the nano-layers. Accordingly, this means cannot be safely extended when a-few-layer structure on a substrate, which is more practical configuration than the suspended periodic multilayered structure, is considered. 

In this work, various a-few-layer structures which are on a substrate is evaluated and the best structure which can have the largest SEW thermal conductivity is discussed. The exact propagation length of the SEWs supported in each configuration is numerically obtained first, and the approximate analytical solution is derived for each configuration. Then, based on this analytical description, the effects of the constituting layers and the configurations on the propagation length of the SEWs are thoroughly explored to achieve a significant enhancement in SEW thermal conductivity. Finally, the configuration is optimized with the genetic algorithm to maximize the SEW thermal conductivity by employing the analytical solutions.

Depending on their nature, the SEWs have been classified into the following modes: surface phonon (plasmon) polaritons, Zenneck modes, and transverse magnetic (TM) guided modes \cite{yang1991long,gluchko2017thermal}. Among these modes, focus usually lied on surface polaritons \cite{raether1988surface} because they can significantly enhance the tunneling of evanescent waves across a vacuum gap and their dispersion relation can be tailored by the structure, and thus, can be employed in controlling near-field thermal radiation \cite{lim2018tailoring,song2016radiative,iizuka2018significant}, near-field thermophotovoltaics \cite{messina2013graphene,lim2018optimization} and microscopy \cite{de2006thermal}. However, a recent experimental study \cite{gluchko2017thermal} showed that the Zenneck modes can provide longer propagation lengths in wider frequency intervals than surface polaritons, which means that Zenneck modes are more beneficial in carrying energy than these latter polaritons. Throughout the manuscript, both Zenneck modes and surface polaritons will be analyzed, while the TM guided modes will be not considered, because they are not coupled to the thermal sources.

\begin{figure}[!t]
\centering\includegraphics[width=0.8\textwidth]{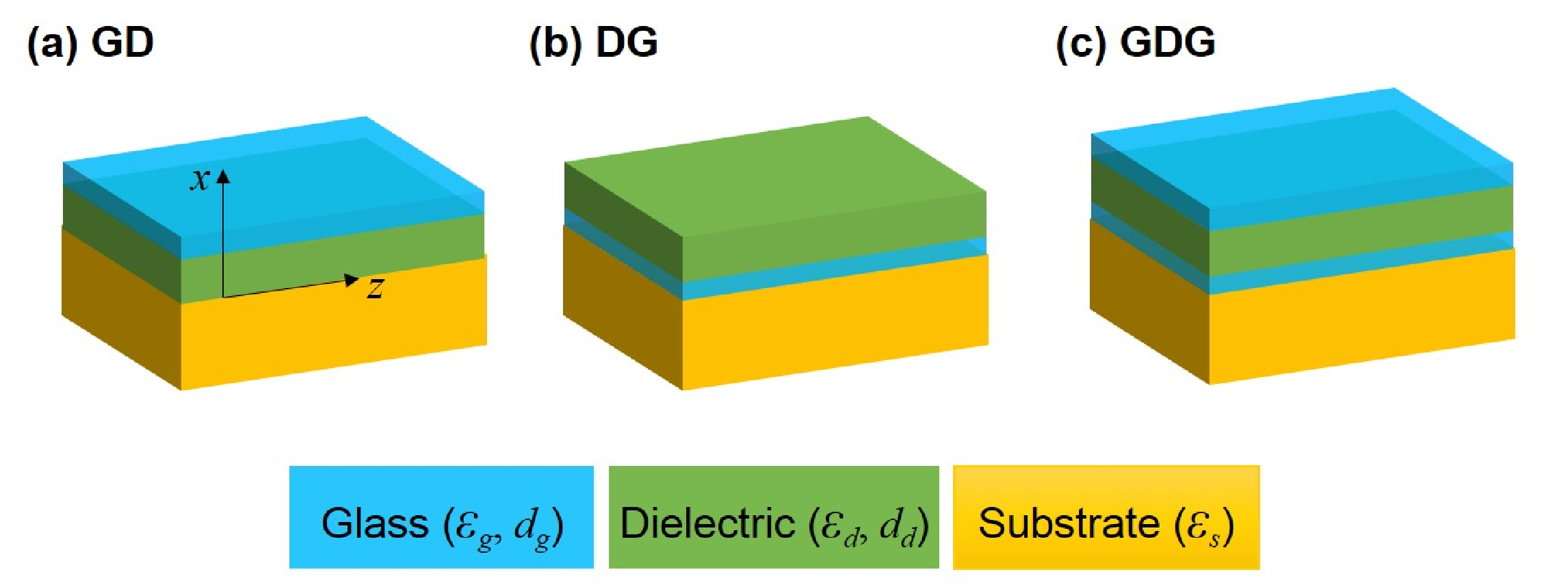}
\caption{Schematics of the layered structures that are considered in this work: (a) GD (glass-dielectric), (b) DG (dielectric-glass), and (C) GDG (glass-dielectric-glass) deposited on a semi-infinite substrate. Volume filling factor of glass is kept constant for the three structures. Subscripts $g$, $d$, and $s$ refer to the glass, dielectric, and substrate, respectively where $\epsilon_{n}$ and $d_{n}$ are the dielectric function and thickness of layer $n$.}
\label{Fig:1}
\end{figure}

\section{Thermal conductivity modeling and dispersion relation}
 The effective thermal conductivity, $k$, of layered structures due to the propagation of SEWs is given by \cite{chen2005surface, ordonez2013anomalous}:
\begin{equation}
k = \frac{1}{4\pi d} \int_0^{\infty}\hbar\omega\Lambda\beta_{R}\frac{\partial f_0}{\partial T} d\omega = \int_0^{\infty} k_\omega d\omega
\end{equation}
where $\hbar$ is the Planck constant divided by $2\pi$,  $\omega$ is the angular frequency, $d$ is the total thickness of the multilayer with average temperature $T$, and $f_0$ is the Bose-Einstein distribution function. 
The values of the unknown parameters $\beta_{R}=\Re(\beta)$ and propagation length $\Lambda = 1/(2\Im(\beta))$ can be obtained respectively from the real and the imaginary parts of the in-plane wavevector $\beta = \beta_{R} + i \beta_{I}$, which is defined by the dispersion relation of the SEWs propagating in a given layered structure. Considering that $\Lambda$ cannot be longer than the system size $L$, throughout the manuscript, the thermal conductivity in Eq.\ (1) is then calculated with $\Lambda_\text{eff}$ which is defined as $1/\Lambda_\text{eff}=1/\Lambda + 1/L$. The frequency interval for integration is taken as that for Zenneck modes and surface phonon polations \cite{gluchko2017thermal}. In Eq.\ (1), the spectral thermal conductivity, $k_\omega$ is proportional to a product of $\hbar\omega\frac{\partial f_0}{\partial T} $ and $\Lambda\beta_{R}$. Given that the value of $\hbar\omega\frac{\partial f_0}{\partial T} $ is larger at lower frequencies, the structure should support SEWs with large propagation length at low frequency to have large thermal conductivities. 


\subsection{Glass-Dielectric structures}
\begin{figure}[!b]
\centering\includegraphics[width=1.0\textwidth]{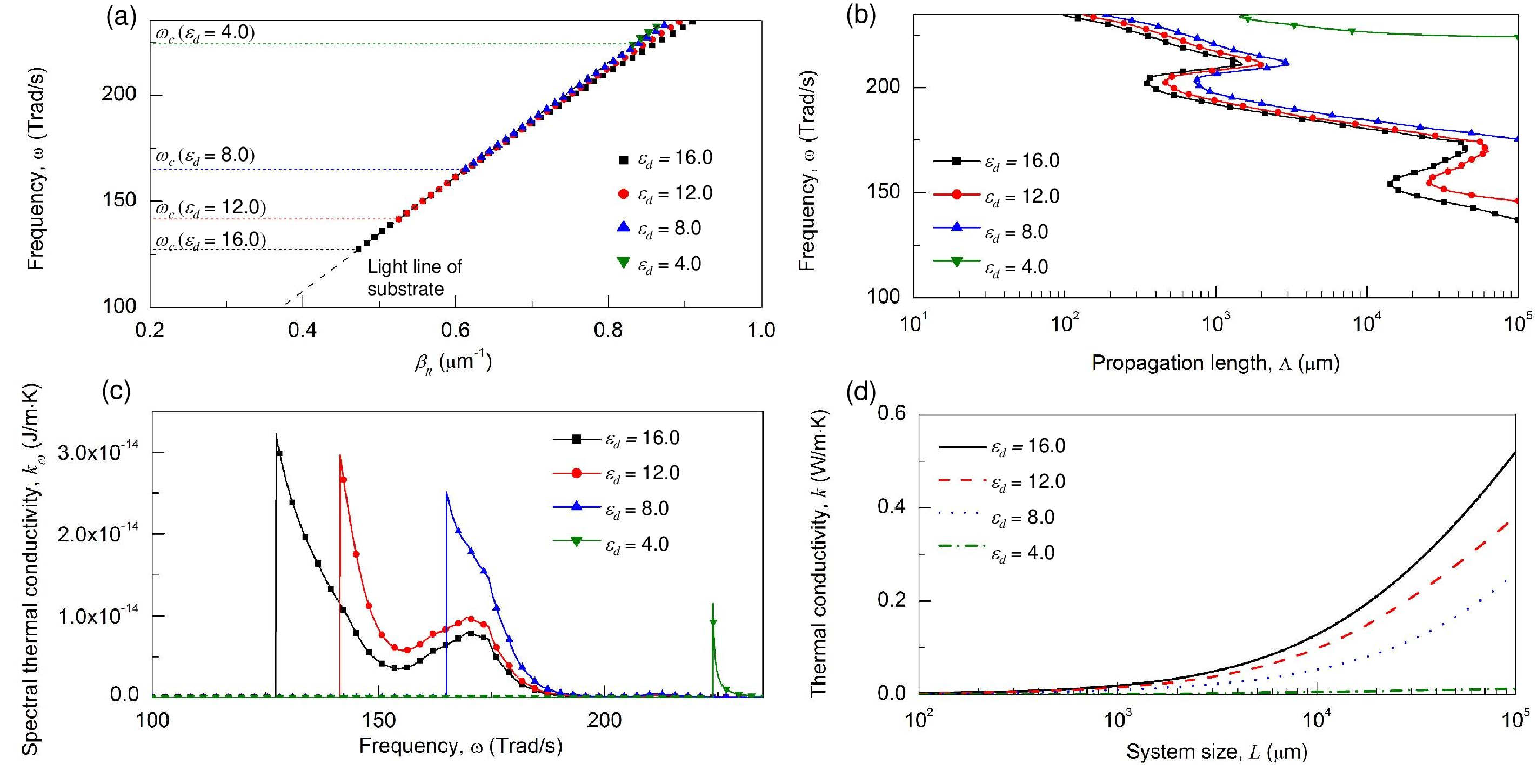}
\caption{Dispersion relation and SEW thermal conductivity of the GD structure with dielectric layers of various permittivities. The thicknesses of the glass and dielectric layers are set to 20 nm and 680 nm, respectively (total thickness $d$ = 700 nm) and average temperature $T$ is set to 300 K. Frequency dependent (a) $\beta_{R}$, (b) propagation length, and (c) spectral thermal conductivity when system size is set to be 10 cm. (d) Resulting thermal conductivity depending on the size of the system.}
\label{Fig:2}
\end{figure}

Let us first consider the Glass-Dielectric (GD) structure (see Fig.\ 1(a)) which supports the propagation of the SEWs with the following dispersion relation \cite{ordonez2014thermal,yeh2008essence}:
\begin{equation}
\frac{\text{tanh}(p_{d} d_{d})+\alpha_{sd}}{1+\alpha_{sd} \text{tanh}(p_{d} d_{d})}
 = -\alpha_{gd} \frac{\text{tanh}(p_{g} d_{g})+\alpha_{0g}}{1+\alpha_{0g} \text{tanh}(p_{g} d_{g})}
\end{equation}
where $\alpha_{ij} = \epsilon_{i} p_{j}$/$\epsilon_{j} p_{i}$. Here, $ \epsilon_{i}$ refers to the dielectric function of each layer and the transverse wavevector $p_{i}$ is given by $p^2_{i} = \beta^2 - \epsilon_{i}k^2_0$ with $k_0=\omega/c_0$ being the wavevector in vacuum. The subscripts 0, $g$, $d$, and $s$ stand for the vacuum, glass, dielectric, and substrate, respectively and $c_0$ is the speed of light in vacuum. The solution of Eq.\ (2) for the complex wavevector $\beta$ is obtained numerically by the secant method with a proper choice of the initial point and using the glass dielectric function reported in \cite{palik1998handbook}. The dielectric function of the substrate is assumed to be $\epsilon_{s}$=1.24 throughout this work, as done in \cite{ordonez2013anomalous, ordonez2014thermal}.

The wavevector $\beta_{R}$ and propagation length of SEWs are shown in Figs.\ 2(a) and 2(b), for different values of the dielectric function, $\epsilon_{d}$. Note that $\beta_{R}$ is close to the light line of substrate (i.e., $\beta_{R} \sim \sqrt{\epsilon_{s}}k_0$) except for the high frequencies ($\omega > $ 200 Trad/s). In this work, this photon-like mode is only considered, because it has much larger propagation lengths than phonon-like mode, which is found to have propagation length smaller than 1 $\mu$m, as in Ref.\ \cite{chen2005surface}. It can be seen that a frequency interval where solution exists, varies with $\epsilon_{d}$. We set the lowest value of frequency, where solution can exist, as a cutoff frequency, $\omega_c$ for each case. Estimating this $\omega_c$ for different configurations is crucial because at $\omega_c$, SEWs with the longest propagation length can be supported in the structures as can be seen in Fig.\ 2(b). When the system size is taken to be 10 cm, the corresponding spectral thermal conductivity, $k_\omega$, for each case is plotted in Fig.\ 2(c), which shows that $k_\omega$ reaches its highest value at $\omega_c$ for each case. Although the propagation length is smaller for higher values of $\epsilon_{d}$ at the same frequency, the wavevector can be extended to lower frequencies with higher values of $\epsilon_{d}$  [refer to Fig.\ 2(b)]. As a result, in Fig.\ 2(d), when $\epsilon_{d} = 16.0$ and system size is set to 10 cm, the corresponding thermal conductivity is 0.52 W/m$\cdot$K, which corresponds to 37$\%$ of the thermal conductivity of bulk glass. Although the solutions shown in Fig.\ 2 can be obtained numerically, it is hard to see which factors affect the propagation length or frequency interval. Alternatively, an analytical solution, which may provide a physical insight on which structures are beneficial for enhancing SEW heat transfer, is derived with several approximations.

Given that the propagation length of SEWs decreases as the thickness of the glass layer increases \cite{chen2005surface, ordonez2013anomalous}, the case of glass with very small thickness will be considered to derive an analytical solution of Eq.\ (2). Under this thin-film approximation (i.e., $\lvert p_{g} \rvert d_{g} << 1$) and up to a first-order on $\lvert p_{g} \rvert d_{g} << 1$,  Eq.\ (2) can be expressed as follows:
\begin{equation}
\frac{ A+(\epsilon_{s} p_{d}/\epsilon_{d} p_{s}) }{1+A (\epsilon_{s} p_{d}/\epsilon_{d} p_{s})}
 = -\frac{\epsilon_{g} p_{d}}{\epsilon_{d} p_{g}}  \frac{p_{g} d_{g}+(\epsilon_0 p_{g}/\epsilon_{g} p_0)}{1+(\epsilon_0 p_{g}/\epsilon_{g} p_0) p_{g} d_{g}}
\end{equation}
where $ A =\text{tanh}(p_{d} d_{d})$. The solution of Eq.\ (3) for $p_{s}$ can be expressed as:
\begin{equation}
p_{s}
 =-\frac{\epsilon_{s} p_{d} [A \epsilon_{g}  (\epsilon_0 + d_{g}\epsilon_{g} p_0)p_{d} + \epsilon_{d} (\epsilon_{g}p_0 + d_{g}\epsilon_0 p^2_{g})] }{\epsilon_{d}[\epsilon_{g}(\epsilon_0 + d_{g}\epsilon_{g}p_0) p_{d} + A \epsilon_{d}(\epsilon_{g}p_0 + d_{g} \epsilon_0 p^2_{g})]}
\end{equation}
Considering that $p_{s} \rightarrow 0$ when $\beta_{R} \rightarrow \sqrt{\epsilon_{s}}k_0$ and $\beta_{I} \rightarrow 0$, $p^2_{i} = p^2_{s} + (\epsilon_{s} - \epsilon_{i})k^2_0$ can be approximate as $p^2_{i} \approx (\epsilon_{s} - \epsilon_{i})k^2_0$. Taking into account the condition $\Re{(p_{i})} >0$ to ensure the bounding of SEW to the interfaces, $p_{i} \approx \sqrt{\epsilon_{s} - \epsilon_{i}}k_0 = D_{i} k_0$. The application of $p_{i} \approx D_{i} k_0$ to Eq.\ (4) provides the approximate solution as
\begin{equation}
p_{s}
 =-\frac{\epsilon_{s} D_{d} k_0 [A' D_{d} \epsilon_{g} (\epsilon_{0}+ D_0 d_{g}\epsilon_{g} k_0)+\epsilon_{d}  (D_0 \epsilon_{g} + d_{g} D^2_{g}\epsilon_0 k_0) ] }{\epsilon_{d}[D_{d}\epsilon_{g}(\epsilon_0 + D_0 d_{g}\epsilon_{g}k_0) + A' \epsilon_{d}(D_0\epsilon_{g} + d_{g} D^2_{g} \epsilon_0 k_0)]}
\end{equation}
where $A' = \text{tanh}(D_{d} k_0 d_{d})$. Figures 3(a) and 3(b) show the analytical solution for $\beta_{R}$ and propagation length, obtained using the relation $\beta^2 = p^2_{s} + \epsilon_{s}k^2_0$. Although several approximations were made, it is readily seen that the approximate solution agrees reasonably with the numerical one. In order to analyze the propagation length and frequency interval where the SEWs can exist, the analytical solution for $\beta_{I}$ is also derived. From the relation 
$\beta^2 = p^2_{s} + \epsilon_{s}k^2_0$, when $\beta_{R} \rightarrow \sqrt{\epsilon_{s}}k_0$, $\beta_{I}$ can be approximated as $(p_{s,R} p_{s,I})/\sqrt{\epsilon_{s}}k_0$ \cite{ordonez2013anomalous}. Here, $p_{s,R}$ and $p_{s,I}$ are the real and imaginary parts of $p_{s}$, respectively. Note that $D_0$ is purely real and $D_d$ is purely imaginary, such that $A' = \text{tanh}(D_{d} k_0 d_{d})$ can be expressed as $A' = i \text{tan}(D_{d,I} k_0 d_{d}) = i A'_I$ with $D_{d} = i D_{d,I}$. Equation (5) can then be re-written as:
\begin{equation}
 \begin{split}
p_{s} &=-\frac{\epsilon_{s} D_{d,I} k_0 [-A'_{I}  D_{d,I} \epsilon_{g} (\epsilon_{0}+ D_0 d_{g}\epsilon_{g} k_0)+\epsilon_{d}  (D_0 \epsilon_{g}+ d_{g} (\epsilon_{s}-\epsilon_{g})\epsilon_0 k_0) ] }{\epsilon_{d}[D_{d,I}\epsilon_{g}(\epsilon_0 + D_0 d_{g}\epsilon_{g}k_0) + A'_{I} \epsilon_{d}(D_0\epsilon_{g} + d_{g} (\epsilon_{s}-\epsilon_{g}) \epsilon_0 k_0)]}
\end{split}
\end{equation}
where $\epsilon_{g} = \epsilon_{g,R} + i \epsilon_{g,I}$. After conducting the complex expansion of Eq.\ (6), $p_{s,R}$ and $p_{s,I}$ can be obtained. Using the relaton $\beta_{I} \approx (p_{s,R} p_{s,I})/\sqrt{\epsilon_{s}}k_0$ and assuming that $p_{g}d_{g}$ is small, the approximate explicit expression for $\beta_{I}$ can be obtained as follows:
\begin{equation}
 \begin{split}
\beta_{I,GD} \approx \frac{D^3_{d,I} d_{g} \epsilon_{g,I} \epsilon^{1.5}_{s}  (1+A'^2_{I})(A'_{I}D_{d,I}\epsilon_0 - D_0 \epsilon_{d}) [D^2_0 ( \epsilon^2_{g,R} +  \epsilon^2_{g,I}) +\epsilon^2_0 \epsilon_{s}] k^2_0}{\epsilon_d (D_{d,I}\epsilon_0 +A'_{I} D_0 \epsilon_{d})^3 ( \epsilon^2_{g,R} +  \epsilon^2_{g,I})}
\end{split}
\end{equation}
\begin{figure}[!t]
\centering\includegraphics[width=0.68\textwidth]{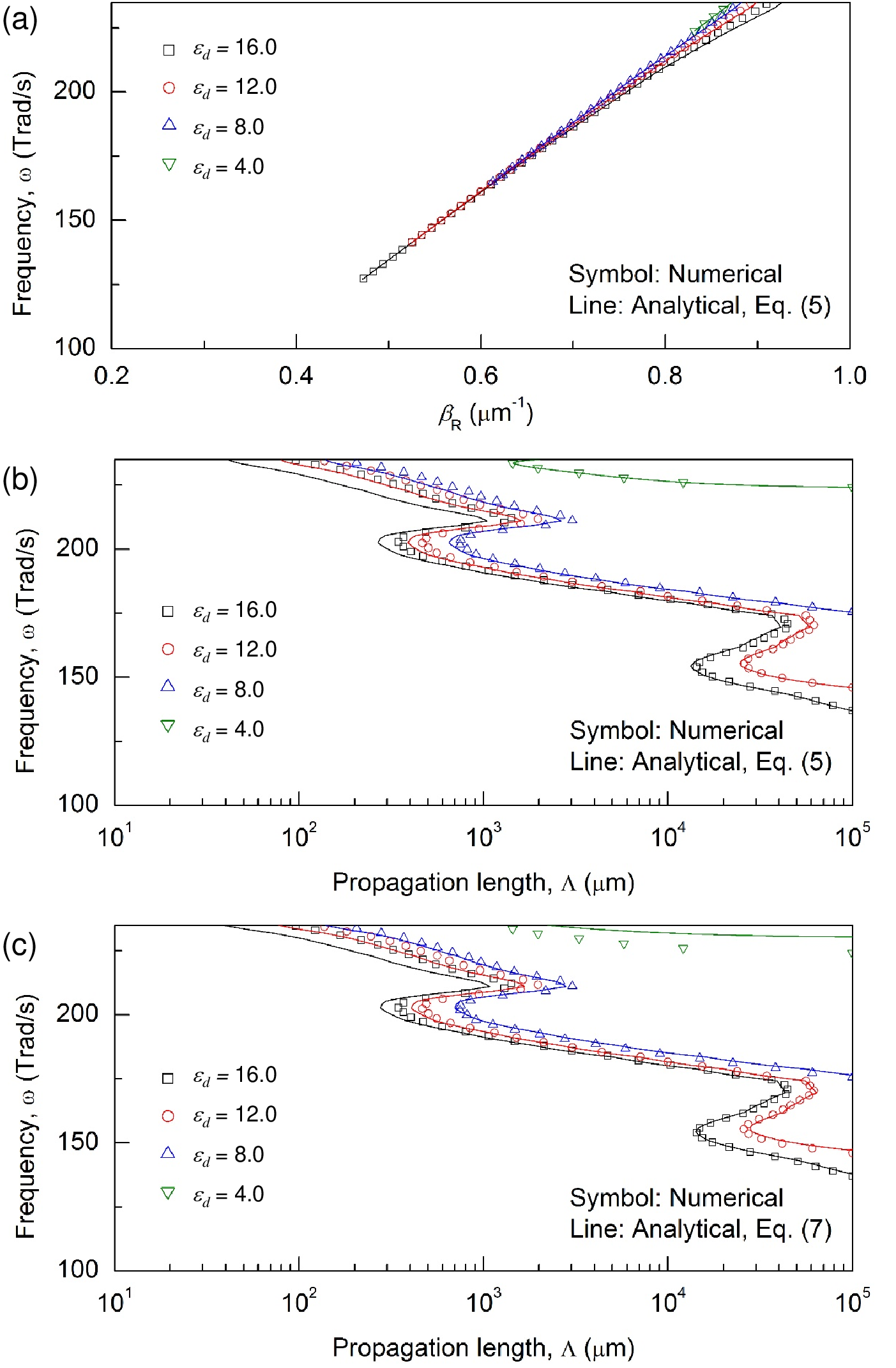}
\caption{ (a) Dispersion relation and (b)-(c) propagation length of SEWs in GD structure obtained via numerical method and analytical solution. Analytical solution is calculated using Eq.\ (5) for (a)-(b) and Eq.\ (7) for (c). The conditions are the same as  in Fig.\ 2. }
\label{Fig:3}
\end{figure}
\begin{figure}[!b]
\centering\includegraphics[width=0.72\textwidth]{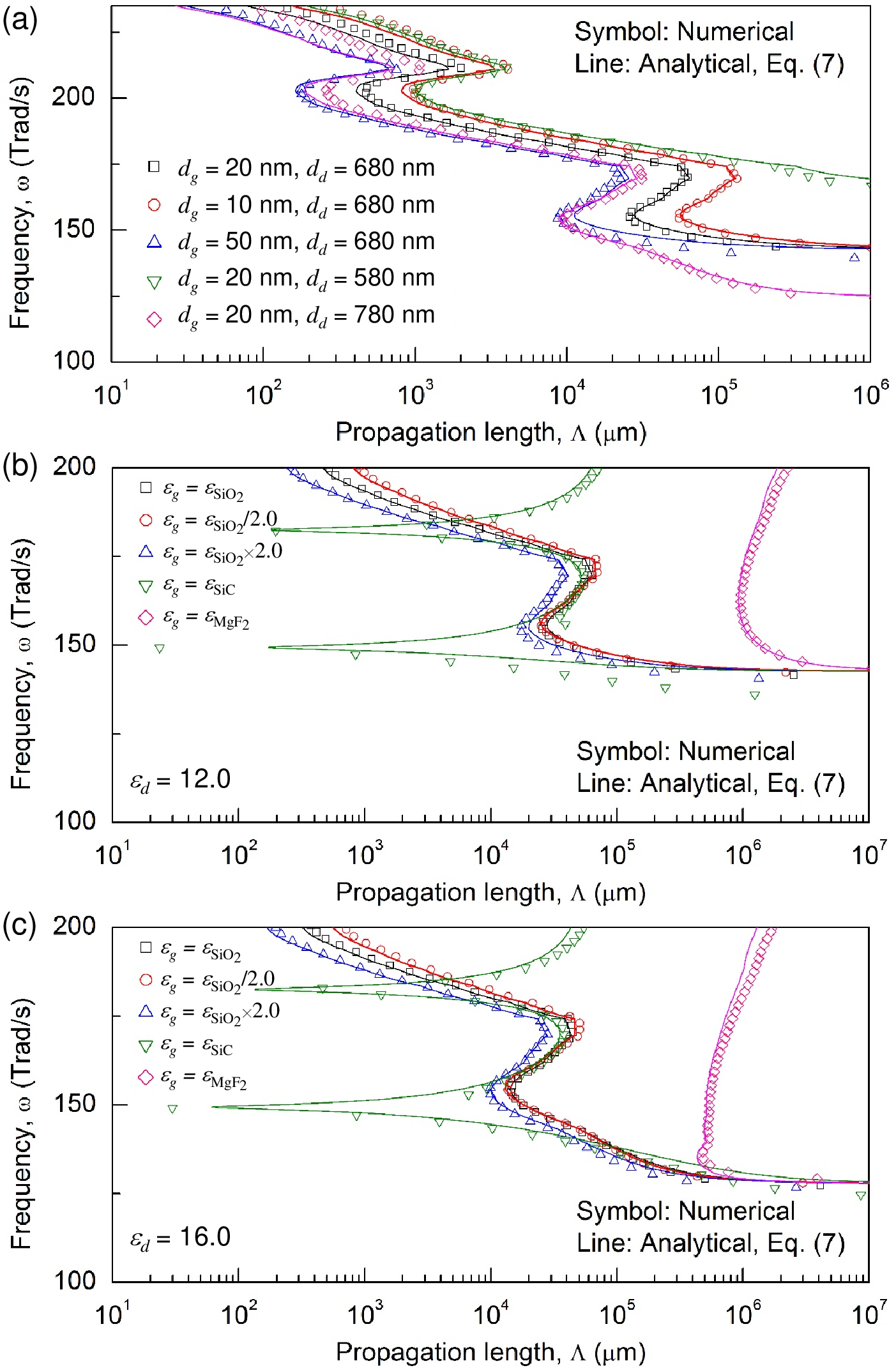}
\caption{The propagation length of SEWs propagating along the GD structure with different configurations, (a) when the thickness of each layer varies while the dielectric function of the dielectric layer $\epsilon_d$ remains as 12.0, (b) and (c) when the dielectric function of the glass layer varies while the thicknesses of glass (SiC or MgF$_2$) layer and dielectric layer are set to 20 nm and 680 nm, respectively. The dielectric function of the dielectric layer is (b) 12.0 and (c) 16.0.}
\label{Fig:4}
\end{figure}
The predictions of Eq.\ (7) for $\beta_{I,GD}$ are shown in Fig.\ 3(c) in comparison with the numerical solution obtained from Eq.\ (2).  Although Eq.\ (7) has been derived through different assumptions, negligible discrepancy between the analytical and numerical solutions is observed except for higher frequency regime where spectral thermal conductivity has smaller values. Taking into account that $D_{d,I}>0$, $d_{g}>0$, $\epsilon_{s} >0$, $\epsilon_{d} >0$, and $\epsilon_{g,I}>0$, $\beta_{I,GD}$ is positive only when $(A'_{I}D_{d,I}\epsilon_0 - D_0 \epsilon_{d})/(D_{d,I}\epsilon_0 +A'_{I} D_0 \epsilon_{d})>0$. In other words, SEWs propagating along the GD structure exist for frequencies satisfying this latter condition, which varies with $\epsilon_{d}$, as can be seen in Fig.\ 3(c). Higher values of $\epsilon_{d}$ satisfy $(A'_{I}D_{d,I}\epsilon_0 - D_0 \epsilon_{d})/(D_{d,I}\epsilon_0 +A'_{I} D_0 \epsilon_{d})>0$ from lower frequency, such that the GD structure with higher $\epsilon_{d}$ can have higher SEW thermal conductivity, as shown in Fig.\ 2(d).

Let us see how the other factors affect the propagation length. If we change $d_{g}$ under the assumption that $p_{g} d_{g}$ is small, as in Fig.\ 4(a), the propagation length increases  as the $d_{g}$ decreases, and the frequency interval where solution exists does not change much by changing $d_{g}$. This trend agrees well with the prediction by Eq.\ (7) as $\beta_{I,GD}$ is proportional to $d_{g}$ and $(A'_{I}D_{d,I}\epsilon_0 - D_0 \epsilon_{d})/(D_{d,I}\epsilon_0 +A'_{I} D_0 \epsilon_{d})$ is not a function of $d_{g}$. Reminding that $A'_I=\text{tan}(D_{d,I} k_0 d_{d})$, the frequency interval where solution exists increases as the $d_{d}$ increases. It should be noted that when $d_{d} = 580$ nm, the solution in Eq.\ (7) exists from relatively high frequencies where the spectral thermal conductivity has small values. It is also worthwhile to mention that for $d_{d} =  380$ nm and $\epsilon_{d} = 12.0$, the solution does not exist within the frequency range where SiO$_2$ can emit. Thus, the proper choice of the dielectric layer thickness is required to exploit the SEWs for carrying energy.

Given that the term $(A'_{I}D_{d,I}\epsilon_0 - D_0 \epsilon_{d})/(D_{d,I}\epsilon_0 +A'_{I} D_0 \epsilon_{d})$ does not involve the dielectric function of glass $\epsilon_{g}$, the frequency interval where the solution exists remains almost invariant under the changes of $\epsilon_{g}$, as can be seen in Figs.\ 4(b) and 4(c). Although a slight mismatch between analytical and numerical solutions is found due to large absolute values of the dielectric function of SiC, $\lvert \epsilon_\text{SiC} \rvert$, the frequency interval does not change much, even if the glass is replaced by SiC or MgF$_2$ in contrast to the notable difference in frequency interval when $\epsilon_{d}$ is varied from 12.0 to 16.0  (see Figs.\ 4(b) and 4(c)).

\subsection{Dielectric-Glass structure}

\begin{figure}[!t]
\centering\includegraphics[width=0.8\textwidth]{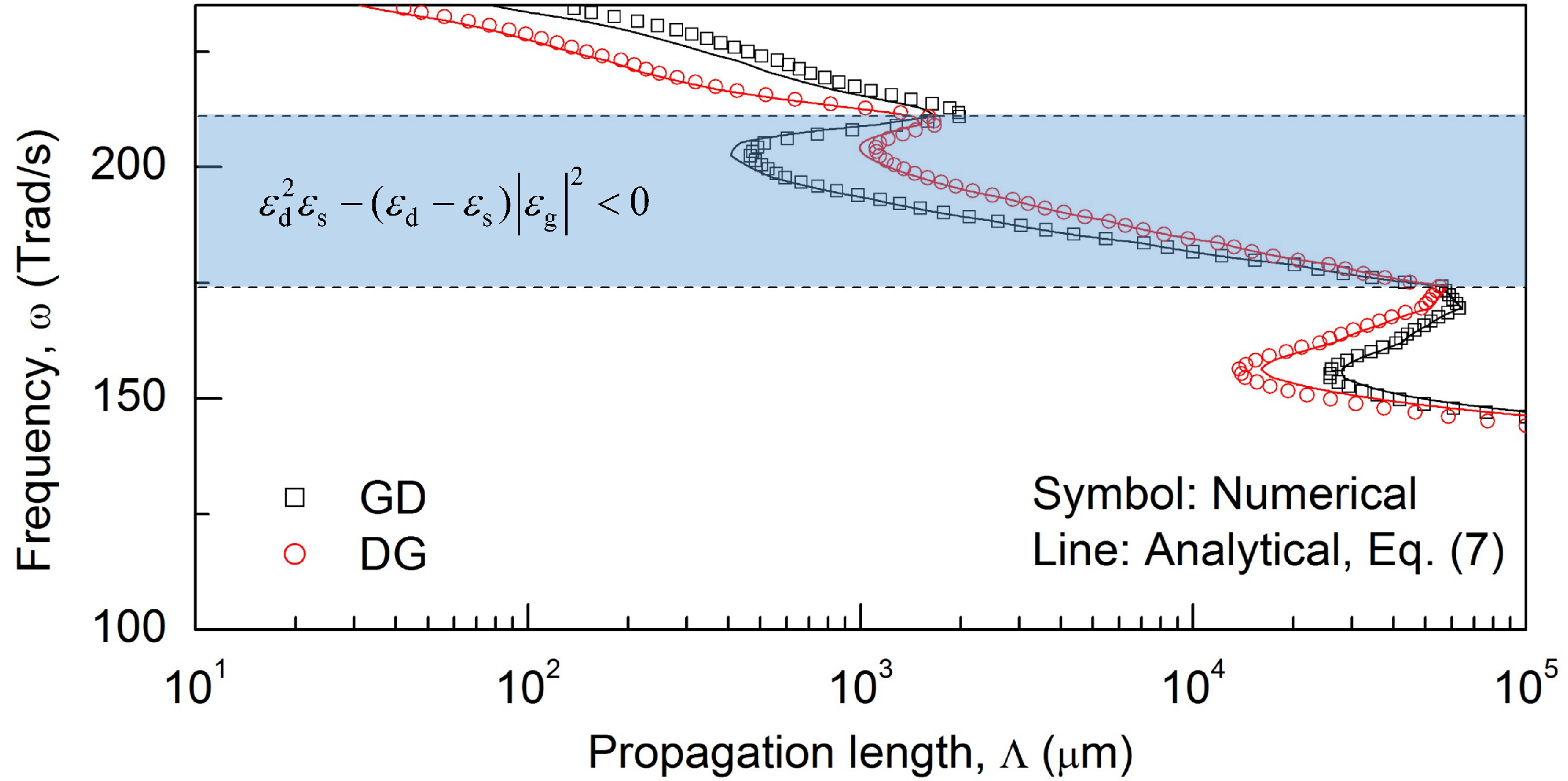}
\caption{Propagation length of the SEWs propagating along the GD and DG structures. Calculations were carried out with thicknesses of the glass and dielectric layers of 20 nm and 680 nm, respectively. The dielectric function of the dielectric layer is set to 12.0.}
\label{Fig:5}
\end{figure}
In this section, the Dielectric-Glass (DG) structure introduced in Fig.\ 1(b) is analyzed to determine its performance for carrying thermal energy with SEWs. As in Eq.\ (2), the dispersion relation for the structure DG can be written as follows \cite{ordonez2014thermal,yeh2008essence}:
\begin{equation}
\frac{\text{tanh}(p_{g} d_{g})+\alpha_{sg}}{1+\alpha_{sg} \text{tanh}(p_{g} d_{g})}
 = -\alpha_{dg} \frac{\text{tanh}(p_{d} d_{d})+\alpha_{0d}}{1+\alpha_{0d} \text{tanh}(p_{d} d_{d})}
\end{equation}
where all the parameters have been previously defined. As in Eqs.\ (3)-(7), the simplification of Eq.\ (8) under the assumption of $\lvert p_{g} \rvert d_{g} << 1$ yields the following $\beta_{I}$ for the structure DG:
\begin{equation}
 \begin{split}
\beta_{I,DG} \approx \frac{D_{d,I} d_{g} \epsilon_{g,I} \epsilon^{1.5}_{s}  (A'_{I}D_{d,I}\epsilon_0 - D_0 \epsilon_{d}) B_{DG} k^2_0}{\epsilon^3_d (D_{d,I}\epsilon_0 +A'_{I} D_0 \epsilon_{d})^3 ( \epsilon^2_{g,R} +  \epsilon^2_{g,I})}
\end{split}
\end{equation}
where
\begin{equation}
\begin{aligned}
B_{DG} = \epsilon^2_{d} (\epsilon_{d}-\epsilon_{s})[D^2_0 ( \epsilon^2_{g,R} +  \epsilon^2_{g,I}) +\epsilon^2_0 \epsilon_{s}] + A'^2_{I}[\epsilon^2_0  ( \epsilon^2_{g,R} +  \epsilon^2_{g,I})  (\epsilon_{d}-\epsilon_{s})^2 + D^2_0 \epsilon^4_{d} \epsilon_{s}] \\+2 A'_{I} D_0 D_{d,I} \epsilon_0 \epsilon_{d} [\epsilon^2_{d} \epsilon_{s} - (\epsilon_{d}-\epsilon_{s})  ( \epsilon^2_{g,R} +  \epsilon^2_{g,I})]
\end{aligned} 
\end{equation}
The difference between $\beta_{I,DG}$ and $\beta_{I,GD}$ can then be expressed as:
\begin{equation}
\begin{aligned}
\beta_{I,DG}- \beta_{I,GD} =\frac{D_{d,I} d_{g} \epsilon_{g,I} \epsilon^{1.5}_{s}  (A'_{I}D_{d,I}\epsilon_0 - D_0 \epsilon_{d}) C k^2_0}{\epsilon^3_d (D_{d,I}\epsilon_0 +A'_{I} D_0 \epsilon_{d})^3 ( \epsilon^2_{g,R} +  \epsilon^2_{g,I})}
\end{aligned} 
\end{equation}
where
\begin{equation}
\begin{aligned}
C= A'_{I}\{2D_0D_{d,I}\epsilon_0 \epsilon_{d}+ A'_{I}[ \epsilon^2_{d} (\epsilon_{s} - \epsilon_0) + \epsilon^2_0 (\epsilon_{s}-\epsilon_{d})]\}[\epsilon^2_{d} \epsilon_{s} - (\epsilon_{d} - \epsilon_{s}) \lvert \epsilon_{g} \rvert^2] 
\end{aligned} 
\end{equation}
Note that the SEWs propagating along the GD structure exists only when $(A'_{I}D_{d,I}\epsilon_0 - D_0 \epsilon_{d})/(D_{d,I}\epsilon_0 +A'_{I} D_0 \epsilon_{d})>0$, as well as when $D_{d,I}>0$, $d_{g}>0$, $\epsilon_{s} >0$, $\epsilon_{d} >0$, and $\epsilon_{g,I}>0$. Accordingly, the sign of $\beta_{I,DG}- \beta_{I,GD}$ is the same as the sign of $C$ described in Eq.\ (12). Thus, $A'_{I}\{2D_0D_{d,I}\epsilon_0 \epsilon_{d}- A'_{I}[ \epsilon^2_{d} (\epsilon_{s} - \epsilon_0) + \epsilon^2_0 (\epsilon_{s}-\epsilon_{d})]\}$ and $\epsilon^2_{d} \epsilon_{s} - (\epsilon_{d} - \epsilon_{s}) \lvert \epsilon_{g} \rvert^2$ can both act as indicators for determining which structure supports SEWs with longer propagation length. In Fig.\ 5, for given configuration, $A'_{I}\{2D_0D_{d,I}\epsilon_0 \epsilon_{d}- A'_{I}[ \epsilon^2_{d} (\epsilon_{s} - \epsilon_0) + \epsilon^2_0 (\epsilon_{s}-\epsilon_{d})]\}$ is found to remain positive in the frequency interval where the solution exists. Therefore, when $\epsilon^2_{d} \epsilon_{s} - (\epsilon_{d} - \epsilon_{s}) \lvert \epsilon_{g} \rvert^2<0$,  $\beta_{I,DG}$ has smaller value than $\beta_{I,GD}$, meaning that $\Lambda_{DG}>\Lambda_{GD}$. Except this frequency interval, it can be readily noted in Fig.\ 5 that the GD structure supports SEWs with longer propagation lengths. For this configuration, because GD structure supports SEWs with longer propagation lengths at lower frequency interval, GD structure can have higher SEW thermal conductivity. However, when it is applied to other materials, it should be noted that depending on the condition where $\epsilon^2_{d} \epsilon_{s} - (\epsilon_{d} - \epsilon_{s}) \lvert \epsilon_{g} \rvert^2>0$ hold, either GD or DG structure can be beneficial in carrying thermal energy.

\subsection{Glass-Dielectric-Glass structures}
\begin{figure}[!b]
\centering\includegraphics[width=0.7\textwidth]{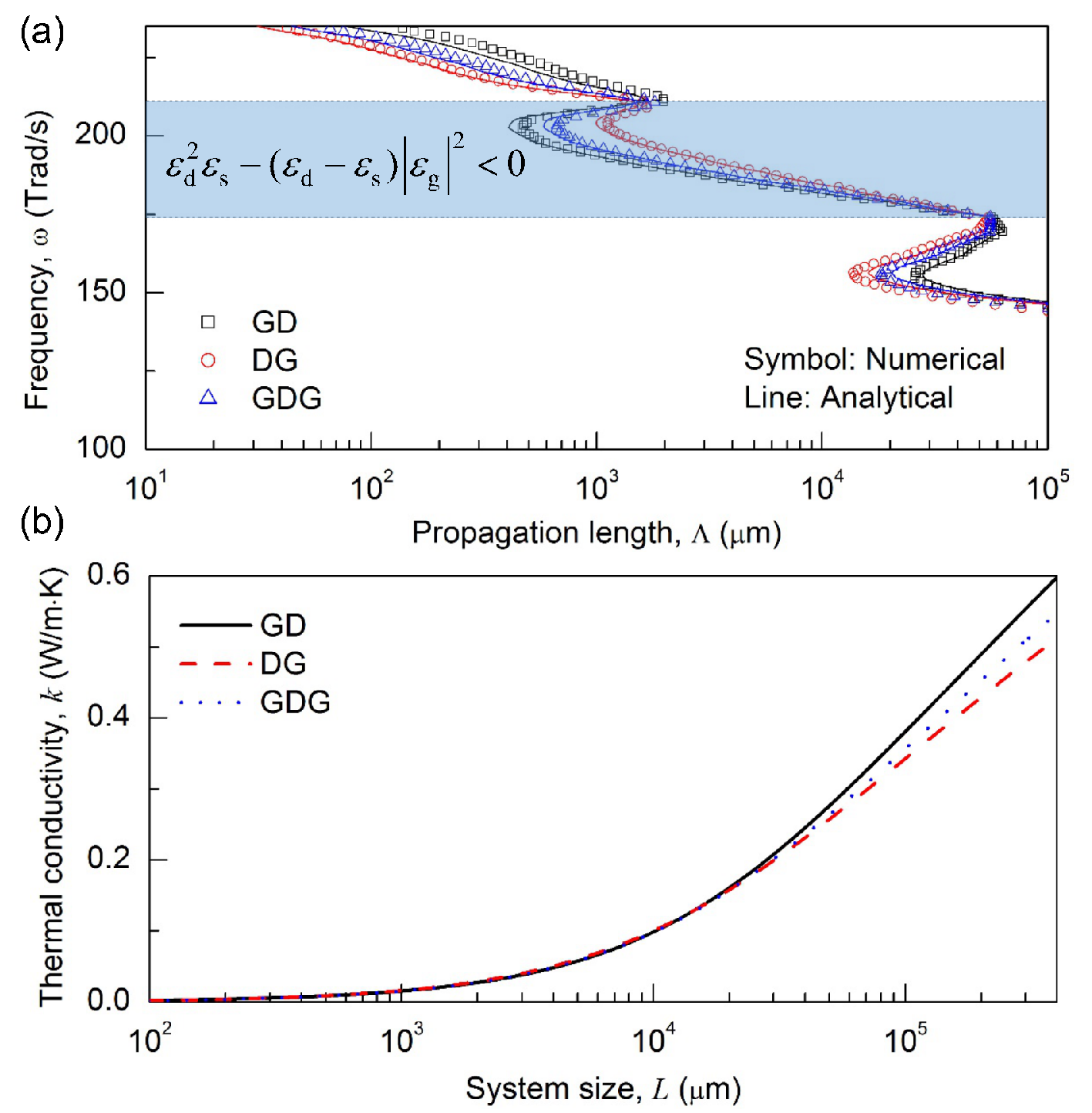}
\caption{(a) Propagation length of the SEWs supported by the GD, DG, and GDG structures. The total thickness of glass layer is set to 20 nm and the thickness of dielectric layer is 680 nm. The dielectric function of the dielectric layer is set to 12.0. (b) Resulting thermal conductivities of each structure when $T =300$ K. }
\label{Fig:6}
\end{figure}
For the three-layer structure shown in Fig.\ 1(c), the following dispersion relation can be derived from the Maxwell equation with proper boundary conditions \cite{yeh2008essence,yeh1977computing}:
\begin{equation}
\begin{aligned}
\text{tanh}(p_{d} d_{d})= -\frac{\alpha_{dg}[1+\alpha_{sg}\text{tanh}(p_{g} \frac{d_{g}}{2})][\alpha_{0g} + \text{tanh}(p_{g} \frac{d_{g}}{2})] + \alpha_{dg}[\alpha_{sg}+\text{tanh}(p_{g} \frac{d_{g}}{2})][1+ \alpha_{0g} \text{tanh}(p_{g} \frac{d_{g}}{2})]}{\alpha^2_{dg}[1+\alpha_{sg}\text{tanh}(p_{g} \frac{d_{g}}{2})][1+\alpha_{0g} \text{tanh}(p_{g} \frac{d_{g}}{2})] +[\alpha_{sg}+\text{tanh}(p_{g} \frac{d_{g}}{2})][\alpha_{0g} + \text{tanh}(p_{g} \frac{d_{g}}{2})] }
\end{aligned} 
\end{equation}
where all the parameters have been previously defined. As in previous sections, an approximate analytical solution for $\beta_{I}$ of SEWs supported in Glass-Dielectric-Glass (GDG) structure can be expressed as:
\begin{equation}
 \begin{split}
\beta_{I,GDG} \approx \frac{D_{d,I} d_{g} \epsilon_{g,I} \epsilon^{1.5}_{s}  (A'_{I}D_{d,I}\epsilon_0 - D_0 \epsilon_{d}) B_{GDG} k^2_0}{2\epsilon^3_d (D_{d,I}\epsilon_0 +A'_{I} D_0 \epsilon_{d})^3 ( \epsilon^2_{g,R} +  \epsilon^2_{g,I})}
\end{split}
\end{equation}
where
\begin{equation}
\begin{aligned}
B_{GDG} = 2\epsilon^2_{d} (\epsilon_{d}-\epsilon_{s})[\epsilon^2_0 \epsilon_{s} + ( \epsilon^2_{g,R} +  \epsilon^2_{g,I})(-\epsilon_0+\epsilon_{s})] \\+ A'^2_{I}(\epsilon_0-\epsilon_{d})[( \epsilon^2_{g,R} +  \epsilon^2_{g,I})  (\epsilon_{d}-\epsilon_{s}) + \epsilon^2_{d} \epsilon_{s}][\epsilon_0 \epsilon_{d} - (\epsilon_0 + \epsilon_{d}) \epsilon_{s}]\\+2 A'_{I} D_0 D_{d,I} \epsilon_0 \epsilon_{d} [\epsilon^2_{d} \epsilon_{s} - (\epsilon_{d}-\epsilon_{s})  ( \epsilon^2_{g,R} +  \epsilon^2_{g,I})]
\end{aligned} 
\end{equation}
In order to compare the SEW thermal conductivity of the GDG structure with those of the GD and DG structures,
$\beta_{I,DG}-\beta_{I,GDG}$ is calculated. Interestingly, the resulting value is found to be half of the value of  $\beta_{I,DG}-\beta_{I,GD}$. 
\begin{equation}
\begin{aligned}
\beta_{I,DG}- \beta_{I,GDG} =\frac{D_{d,I} d_{g} \epsilon_{g,I} \epsilon^{1.5}_{s}  (A'_{I}D_{d,I}\epsilon_0 - D_0 \epsilon_{d}) C k^2_0}{2\epsilon^3_d (D_{d,I}\epsilon_0 +A'_{I} D_0 \epsilon_{d})^3 ( \epsilon^2_{g,R} +  \epsilon^2_{g,I})} \\= \frac{\beta_{I,DG}- \beta_{I,GD}}{2}
\end{aligned} 
\end{equation}
Accordingly, $\beta_{I,DG}- \beta_{I,GDG}$ = $\beta_{I,GDG}- \beta_{I,GD}$. Furthermore, both $\beta_{I,DG}- \beta_{I,GDG}$ and $\beta_{I,GDG}- \beta_{I,GD}$ have positive values when $\epsilon^2_{d} \epsilon_{s} - (\epsilon_{d} - \epsilon_{s}) \lvert \epsilon_{g} \rvert^2>0$ as the  $\beta_{I,DG}- \beta_{I,GD}$ does. When $\epsilon^2_{d} \epsilon_{s} - (\epsilon_{d} - \epsilon_{s}) \lvert \epsilon_{g} \rvert^2>0$, it is readily observed in Fig.\ 6(b) that $\Lambda_{GD}>\Lambda_{GDG}>\Lambda_{DG}$ and vice versa. Among the three structures, the GD one shows the highest SEW conductivity when the maximum system size is set to be 10 cm, because it supports SEWs with the longest propagation length at the lowest frequency regime. 

\section{Discussion}
\begin{figure}[!t]
\centering\includegraphics[width=0.7\textwidth]{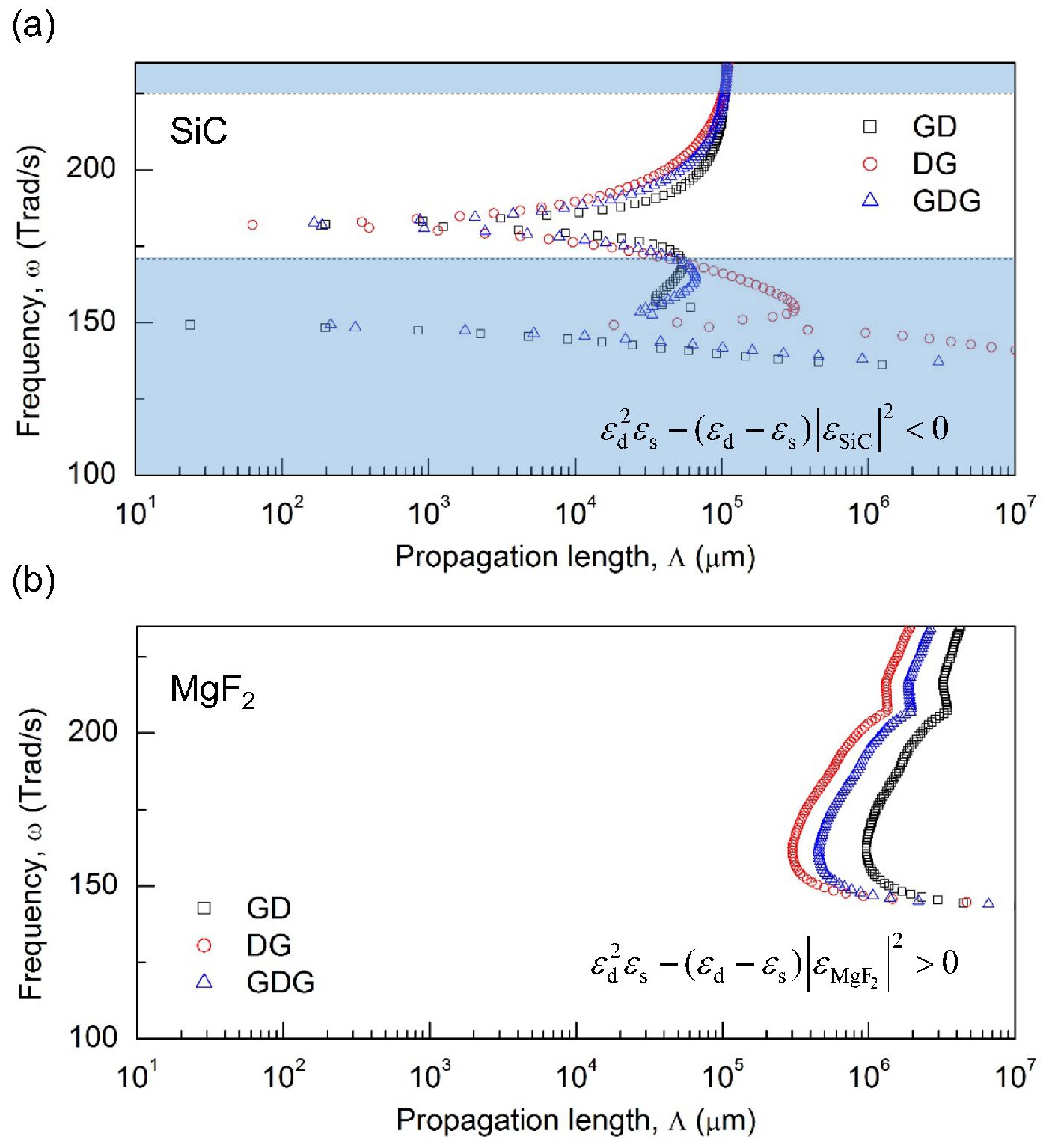}
\caption{ Propagation length of the SEWs supported in GD, DG, and GDG structures. The total thickness of the SiC/MgF$_2$ layer is 20 nm and the thickness of the dielectric layer is 680 nm. The dielectric function of the dielectric layer is set to 12.0 (a) Glass is replaced by SiC. (b) Glass is replaced by MgF$_2$.}
\label{Fig:7}
\end{figure}
As discussed in the previous section, $\epsilon^2_{d} \epsilon_{s} - (\epsilon_{d} - \epsilon_{s}) \lvert \epsilon_{g} \rvert^2$ can act as the indicator for determining the best structure for carrying thermal energy with SEWs. The expression $\epsilon^2_{d} \epsilon_{s} - (\epsilon_{d} - \epsilon_{s}) \lvert \epsilon_{g} \rvert^2>0$ can be re-written as  $\lvert \epsilon_{g} \rvert^2<\frac{\epsilon^2_{d} \epsilon_{s}}{\epsilon_{d} - \epsilon_{s}}$ when $\epsilon_{d} - \epsilon_{s}>0$. Therefore, we can infer that if the glass layer is replaced by a material which has a dielectric function with small absolute value, the propagation length for the GD structure will be the longest among three structures; otherwise, the SEWs for DG structure will have the longest. In Fig.\ 7, the propagation lengths for GD, DG, and GDG structures are shown, when glass  layers are replaced by SiC and MgF$_2$. As predicted, for SiC having a higher dielectric function than glass at lower frequency regime, DG structure shows the longest propagation length at the lowest frequency where SEWs can exist. On the other hand, for MgF$_2$, $\epsilon^2_{d} \epsilon_{s} - (\epsilon_{d} - \epsilon_{s}) \lvert \epsilon_\text{MgF$_2$} \rvert^2>0$ holds for the entire frequency interval, and thus, $\Lambda_{GD}>\Lambda_{GDG}>\Lambda_{DG}$ can be observed for the frequency interval where the SEWs exist. Although we have simplified the analytical solutions for various structures, those solutions can be applied to a wide range of materials and they enable us to predict the structure that holds SEWs with the longest propagation lengths.
 
In addition, it is worthwhile to mention that calculation of SEW thermal conductivity with the analytical solution requires significantly reduced calculation time compared to that for numerical solution. Accordingly, the optimal configurations for maximizing SEW thermal conductivity can be found by employing genetic algorithm. As in Table 1, the maximum calculated SEW thermal conductivity of 0.974 W/m$\cdot$K, which is 70\% of the  thermal conductivity of bulk glass, is obtained with GD structure with $\epsilon_{s}$ of 1.24. For that configuration, the numerical exact solution results in SEW thermal conductivity of 1.00 W/m$\cdot$K. In other words, the maximum thermal conductivity can be obtained with error less than 3\% in much shorter calculation time by employing analytical solution. While varying the range for dielectric function of dielectric layer and substrate and maximum system size, the error is kept less than 5\%. The analytical solution enables us to conduct multiple optimization processes under various boundary conditions with high accuracy. As a result, in Table 1, it is readily noted that thinner glass layer, larger dielectric function of dielectric layers, and smaller dielectric function of substrate are beneficial for enhancing SEW thermal conductivity.

\begin{table}[!t]
\caption{\label{Table1} Bounds for the variables and corresponding optimal configuration and SEW thermal conductivity for GD structure at $T=300$ K} 
\footnotesize
\centering
\begin{tabular}{ >{\centering}p{0.15\textwidth} >{\centering}p{0.15\textwidth}>{\centering}p{0.15\textwidth}>{\centering}p{0.2\textwidth}>{\centering}p{0.2\textwidth}}  
\hline\hline
Variable &Lower bound&Upper bound & Optimal configuration & SEW thermal conductivity [error] (W/m$\cdot$K) \tabularnewline
\hline
$d_{g}$ (nm) & 20   &1000  & 20    \tabularnewline
$d_{d}$ (nm) & 20   &1000  & 695 & 0.512 [1.8\%] \tabularnewline
$\epsilon_{d}$  & 1   &16 & 16  \tabularnewline
$\epsilon_{s}$&   - & - & set: 1.24\tabularnewline \hline

$d_{g}$ (nm) & 5   &1000  & 5    \tabularnewline
$d_{d}$ (nm) & 5   &1000  & 705 & 0.974 [2.7\%] \tabularnewline
$\epsilon_{d}$  & 1   &16 & 16  \tabularnewline
$\epsilon_{s}$&   - & - & set: 1.24\tabularnewline \hline

$d_{g}$ (nm) & 5   &1000  & 5    \tabularnewline
$d_{d}$ (nm) & 20   &1000  & 600 & 1.27  [4.5\%]\tabularnewline
$\epsilon_{d}$  & 1   &16 & 16  \tabularnewline
$\epsilon_{s}$&   1.1 & 16 & 1.1\tabularnewline \hline\hline
\end{tabular}
\end{table}

\section{Conclusions}

In this work, the dispersion relations of SEWs propagating along the GD, DG, and GDG structures have been derived and analyzed along with their corresponding analytical solutions for the in-plane wavevector. The impact of various factors affecting the propagation length and thermal conductivity has been discussed based on the analytical solutions. It has been found that the frequency interval where the SEWs exist is almost independent on the dielectric function of the glass layer. Furthermore, the SEW propagation length of each structure has been compared at different conditions and the indicator for determining the structure where SEWs with the longest propagation length can be supported has been found. The propagation lengths of structure GDG is found to be between those of the structures DG and GD. The optimization to get maximum SEW thermal conductivity is also conducted with genetic algorithm, yielding SEW thermal conductivity of 1.27 W/m$\cdot$K similar to the thermal conductivity of bulk glass. It is worthwhile to mention that thermal energy transport by SEWs has a different nature than conventional thermal conduction; that is, its relatively long propagation length indicates that the energy is weakly absorbed by the substrate. Accordingly, it is a quite effective channel to spread heat from hot spots. Because Si, which is the most widely used material in semiconductor industry, has a dielectric function around 11.7 at 100 Trad/s and is almost lossless \cite{palik1998handbook}, the analysis conducted in this work provides guidelines for enhancing SEW thermal conductivity of nano-sized glass layers combined with Si ones, mitigating the hot-spot issue in electronic devices.

\begin{acknowledgments}
This research was supported by the Basic Science Research Program (Grants No. NRF-2017R1A2B2011192 and NRF-2019R1A2C2003605) through the National Research Foundation of Korea (NRF) funded by Ministry of Science and ICT. The stay of M. Lim in CentraleSupélec has been supported by the Erasmus Mundus EASED programme (Grant 2012-5538/004-001) coordinated by CentraleSupélec.
\end{acknowledgments}

\bibliography{Lim_Bib}

\end{document}